\documentclass[twocolumn]{jpsj2} 
\title{$^{13}$C NMR Study on Zero-Gap State in the Organic Conductor 
$\theta$-(BEDT-TTF)$_{2}$I$_{3}$ under Pressure}

\author{\textsc{Kazuya Miyagawa, Motoaki Hirayama, Masafumi Tamura$^{1}$, 
and Kazushi Kanoda}} 

\inst{Department of Applied Physics, University of Tokyo, Bunkyo-ku, Tokyo 113-8656, Japan \\
$^{1}$ Department of Physics, Faculty of Science and Technology, Tokyo University of Science, 
Noda, Chiba 278-8510, Japan \\}

\abst{We present the results of our $^{13}$C NMR study of the quasi-two-dimensional organic conductor 
$\theta$-(BEDT-TTF)$_2$I$_{3}$ under pressure, which is suggested to be a zero-gap conductor by 
transport measurements. 
	We found that NMR spin shift is proportional to $T$ and that spin-lattice relaxation rate follows 
the power law $T^{\alpha}$ ($\alpha=3 \sim 4$), where $T$ is the temperature.
	This behavior is consistent with the cone-like band dispersion and provides 
microscopic evidence for the realization of the zero-gap state 
in the present material under pressure.}

\kword{zero-gap conductor, NMR, $\theta$-(BEDT-TTF)$_{2}$I$_{3}$, pressure}

\begin{document}
\maketitle

	Graphene provides a unique two-dimensional electron system with a cone-shaped dispersion 
relation, where  quasi-particles are massless Dirac fermions and show an anomalous quantum 
magneto-transport such as an unconventional quantum Hall effect \cite{Ref1_Novoselov}.
	The quasi-two-dimensional organic conductor $\alpha$-(BEDT-TTF)$_{2}$I$_{3}$ ($\alpha$-I$_{3}$) 
under pressure is a candidate bulk version of purely two-dimensional graphene. 
	This material is a charge-ordered insulator below 135 K at ambient pressure \cite{Ref2_Kino}.
	However, Tajima \textit{et al.} found that the system shows temperature-insensitive 
resistivity at high pressures above 15 kbar \cite{Ref3_Tajima}.
	Their subsequent measurements of Hall coefficient indicates its strong temperature dependence 
over several decades from room temperature down to 1 K \cite{Ref3_Tajima}.
	Based on these two transport characteristics, they argued that carrier density decreases to 
a vanishingly small value and that mobility conversely increases to a high value on order of 10$^{5}$ cm$^{2}$/Vs 
with power-law temperature dependences \cite{Ref3_Tajima}.
	The band structure of $\alpha$-I$_{3}$ was first calculated by the 
tight-binding method for the extended H\"{u}kel orbital by Kobayashi \textit{et al.} and shown to be 
semi-metallic with hole and electron pockets at ambient pressure \cite{Ref4_Kobayashi}.
	Recent calculations by Ishibashi \textit{et al.} \cite{Ref5_Ishibashi}, Kino and Miyazaki, \cite{Ref6_Kino} and 
Katayama \textit{et al.} \cite{Ref7_Katayama} have shown that the variation of transfer integrals and/or 
inter-site Coulomb interaction with pressure gives a corn-like band-structure and explains 
the above-mentioned properties of $\alpha$-I$_{3}$ at high pressures.

\begin{figure}
\begin{center}
\includegraphics[width=8.5cm]{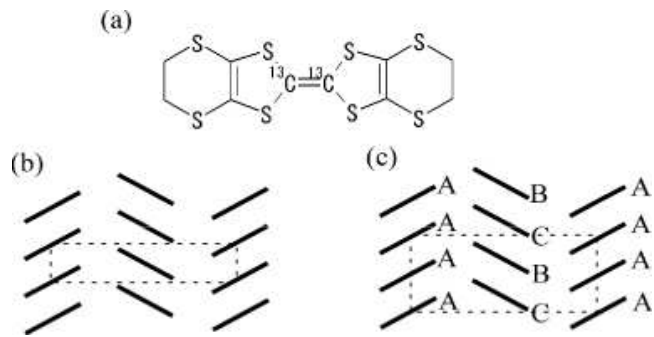}
\caption{\label{Fig1} (a) $^{13}$C enriched BEDT-TTF molecule. In-plane BEDT-TTF molecular 
arrangements in (b) $\theta$-I$_{3}$ (the so-called average structure, where all molecules are 
equivalent) and (c) the $\alpha$-I$_{3}$ (three BEDT-TTF sites labeled $A$, $B$, and $C$ 
in the unit cell are inequivalent).}
\end{center}
\end{figure}

	The same transport characteristics as those of $\alpha$-I$_{3}$ were found in  the analogous compound 
$\theta$-(BEDT-TTF)$_{2}$I$_{3}$ ($\theta$-I$_{3}$) with a similar structure to 
$\alpha$-I$_{3}$ (Figs. \ref{Fig1}(b) and \ref{Fig1}(c)) \cite{Ref8_Tamura,Ref9_Tajima}.
	At ambient pressure, $\theta$-I$_{3}$ is a metal \cite{Ref10_Kobayashi} in constant to $\alpha$-I$_{3}$, 
which is a charge-ordered insulator \cite{Ref2_Kino}.
	At pressures above 5 kbar, however, $\theta$-I$_{3}$ shows 
temperature-insensitive resistivity \cite{Ref8_Tamura, Ref9_Tajima}  and 
temperature-sensitive Hall coefficient \cite{Ref9_Tajima}.
	Thus, there have been reports on an increasing number of experiments and theoretical works that support the notion 
that the organic compound is a bulky material with massless Dirac fermions.
	However, experimental studies have been restricted to transport measurements to date.

	In the present work, we aim at clarifying a possible graphene-like state in 
a bulky organic material by the method of nuclear magnetic resonance (NMR). 
	As a sample, $\theta$-I$_{3}$ is adopted because it has the advantage over $\alpha$-I$_{3}$ that experiments can be carried out at 
much lower pressures.

	Here, we perform a $^{13}$C NMR experiment on $\theta$-I$_{3}$ 
under pressure to examine the possible bulky zero-gap state from the magnetic point of view 
for the first time.
%

	In the present NMR study, we used a $\theta$-I$_{3}$ single crystal where the two central carbon 
sites connected with a double bond in the BEDT-TTF molecule are substituted with a $^{13}$C isotope 
(Fig. \ref{Fig1}(a)).
	The central carbon sites are located in the region of high density of the highest occupied molecular orbital (HOMO). 
	In accordance with the transport results suggesting that the zero gap state is realized above 5 kbar 
\cite{Ref8_Tamura, Ref9_Tajima}, we applied a hydrostatic pressure of 8 kbar to the sample using a NiCrAl-BeCu 
hybrid piston cylinder cell with  a pressure-transmitting medium oil, Daphne 7373.
	We mounted the sample in a NMR coil and applied a magnetic field of 8 Tesla 
parallel to the conducting plane to eliminate the magneto-orbital effect \cite{Ref11_Kobayashi}, 
which is interesting but is out of the scope of the present study.
	It is noted that the alignment is not perfect; therefore, the sample may be slightly deviated 
from the parallel condition.
	$^{13}$C NMR measurement was carried out by the spin-echo method, where the typical width of 
the $\pi /2$ pulse was 2 $\mu$s.
	The NMR spectra were obtained by the first Fourier transformation (FFT) of spin echoes.
	The $^{13}$C NMR frequency of tetramethylsilane (TMS) was used as the 
origin of the NMR shift.
	The nuclear spin-lattice relaxation rate 1/$T_{1}$ was obtained by the standard saturation 
recovery method.
	
	Relaxation curves of $^{13}$C nuclear magnetization were fitted to single exponentials 
for all the temperatures measured.

\begin{figure}
\begin{center}
\includegraphics[width=8.5cm]{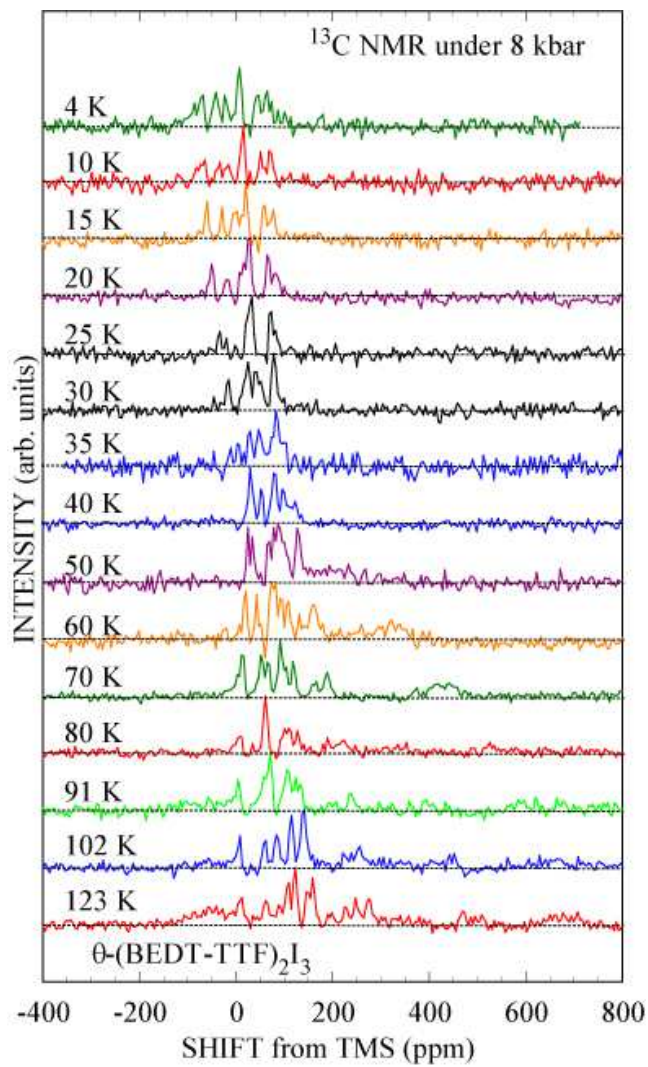}
\caption{\label{Fig2} (Color online) Temperature dependence of $^{13}$C NMR spectra.}
\end{center}
\end{figure}

	The temperature dependence of $^{13}$C NMR spectra is shown in Fig. \ref{Fig2}.
	The line at approximately  -60 ppm at 123 K comes from the pressure medium oil.
	This line broadens at low temperatures owing to the sodification of oil and 
becomes unobservable below 100 K because the relaxation time $T_{1}$ is much longer 
than that of the sample.
	The NMR spectrum of the sample has a rather complicated shape, which consists of many peaks.
	This spectral feature is different from the simple four-line structure observed 
at ambient pressure \cite{Ref12_Hirata}, where the system is an ordinary metal with cylindrical 
Fermi surfaces, as evidenced by \textit{de Haas-van Alphen} oscillations \cite{Ref13_Tokumoto}.
	The four lines are what is expected in the so-called average structure of $\theta$-I$_{3}$ 
(Fig. \ref{Fig1}(b)) determined by XRD analysis \cite{Ref14_Kobayashi} at ambient pressure; 
the two (crystallographycally equivalent) BEDT-TTF molecules in the in-plane unit cell are 
inequivalent with respect to the magnetic field parallel to the conducting 
layer. 
	Within the molecule, two neighboring $^{13}$C atoms, which are equivalent in $\theta$-I$_{3}$, give 
the so-called Pake doublet due to the nuclear dipolar splitting in the BEDT-TTF molecule.
	Thus, a total of four lines are expected.
	The complicated lines show that the structure at 8 kbar is changed from 
that at ambient pressure.
	Although the crystal structure under pressure is not known, the pressure dependence of resistivity 
at room temperature shows a discontinuous jump at approximately 5 kbar \cite{Ref8_Tamura, Ref9_Tajima} 
above which the transport behavior of the zero-gap conductor, as observed in 
$\alpha$-I$_{3}$, appears instead of the metallic behavior at ambient pressure.
	Therefore, we speculate that $\theta$-I$_{3}$ undergoes structural transformation to 
a phase similar to $\alpha$-I$_{3}$, which has a similar 
fish bone like molecular arrangement.
	If the $\alpha$-type arrangement appears above 5 kbar, the unit cell should contain three 
kinds of non-equivalent BEDT-TTF molecules (Fig. \ref{Fig1}(c)).
	They are specified by two $A$ molecules, and one each of the $B$ and $C$ molecules, which form two types 
of columns.
	It is well known that the nuclear dipole-dipole interaction ($^{13}$C=$^{13}$C) brings a maximum of 
4 lines in the case that the neighboring $^{13}$C atoms are inequivalent as in the $A$ molecule of 
$\alpha$-I$_{3}$. 
	Thus, the complicated NMR line shape can be explained by the non-equivalent circumstances of 
BEDT-TTF molecules like the $\alpha$-type arrangement and dipole-dipole interactions between 
$^{13}$C nuclei in BEDT-TTF.

\begin{figure}
\begin{center}
\includegraphics[width=8.5cm]{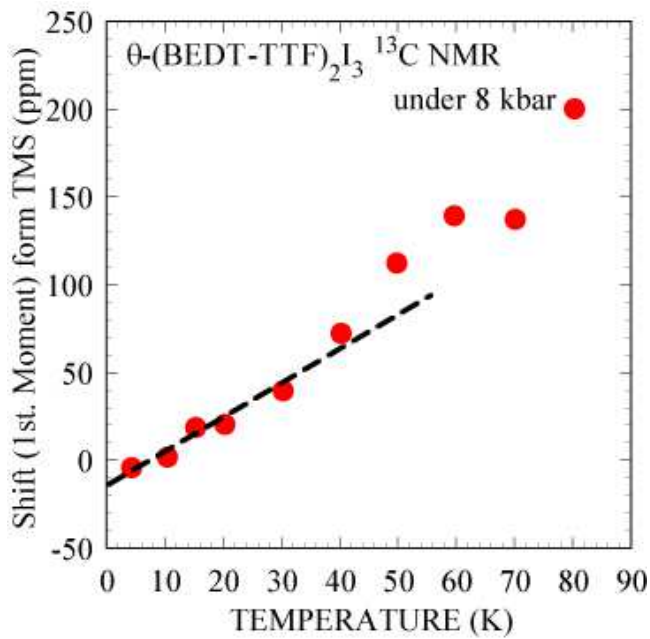}
\caption{\label{Fig3} (Color online) Temperature dependence of $^{13}$C NMR shift. 
	The scale of the shift is relative to the line position of TMS.}
\end{center}
\end{figure}

	With decreasing temperature, the NMR spectra shift to a lower frequency and the distribution of 
resonance frequency becomes smaller, which indicates a decrease in spin susceptibility.
	Figure \ref{Fig3} shows the temperature dependence of the gravity of NMR spectra, 
that is, an average shift.
	It monotonically decreases below 60 K and is proportional 
to temperature  below 40 K .
	The linear extrapolation of the observed shift to 0 K yields -13 ppm. 
	It is interesting that the linearity holds down to the lowest temperature measured of 4 K.
	In the BEDT-TTF molecule, the orbital angular momentum is quenched because of its low structural 
symmetry; so the orbital component of the shift is negligible. 
	In addition, although a possible large diamagnetism expected from the inter band effect 
in the zero-gap state has been theoretically pointed out, \cite{Ref11_Kobayashi} its contribution to the line 
shift should be very small in the parallel-field configuration.
	Thus, the observed line shift is considered to be the sum of a temperature-independent chemical 
shift and a linearly temperature-dependent spin shift.

\begin{figure}
\begin{center}
\includegraphics[width=8.5cm]{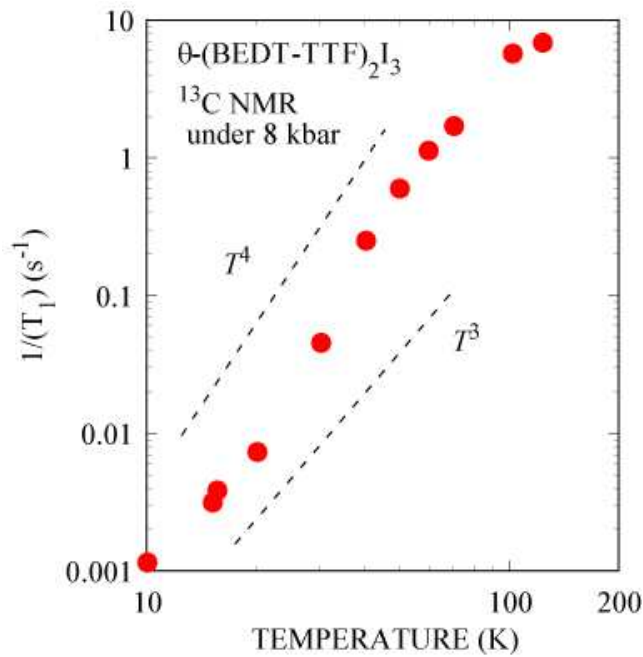}
\caption{\label{Fig4} (Color online) $^{13}$C nuclear spin-lattice relaxation rate.}
\end{center}
\end{figure}

	The temperature dependence of $1/T_{1}$ is shown in Fig. \ref{Fig4}. 
	$T_{1}$ was determined from the recovery curve of the entire spectra, which was well 
approximated to a single exponential function up to 90$\%$ recovery.
	Because the multiple-peak structure in the spectra indicates the presence of different 
hyperfine tensors, one expects a non-single exponential recovery curve.
	A possible explanation for the observed single-exponential nature is in terms of the $T_{2}$ process; 
when $T_{1}$ is much longer than $T_{2}$, the nuclear spin temperatures at different 
$^{13}$C sites reach equilibrium much faster than the nuclear spin-lattice relaxation time.
	As seen in the figure, $1/T_{1}$ shows a steep decrease down to 10 K.
	The temperature dependence seems to be characterized by a power law of $T^{3}$ below 30 K with 
an upward deviation at higher temperatures. 
	An exponential function fails to fit to the data.

	It is evident that these behaviors of the line shift and $1/T_{1}$ are not expected in 
conventional metals.

	It is well known that in a simple metallic case, the spin shift and $1/(T_{1}T)$ are 
proportional to the density of states averaged by temperature, 
$\langle D(\epsilon)\rangle_{kT}$ and $\langle D(\epsilon)\rangle_{kT}^{2}$, respectively, 
where $\epsilon$ is the energy measured from the Fermi energy, $\epsilon_{\rm F}$.
	Because $\langle D(\epsilon)\rangle_{kT}$ is essentially temperature independent 
in a simple metal, the spin shift and $1/(T_{1}T)$ would be temperature-independent.
	The qualitatively different behaviors observed here are explained in terms of the cone-shaped 
band dispersion (Dirac cone) with a Fermi level at the apex.
	The low-energy quasi-particle excitations around the apex of the Dirac cone, which governs 
the magnetic properties, have $D(\epsilon)$ with a linear $\epsilon$ dependence. 
	In this case, the thermal average of the density of states, $\langle D(\epsilon)\rangle_{kT}$, 
is proportional to temperature.
	This leads to the following relations: spin shift $\propto$ $\langle 
D(\epsilon)\rangle_{kT}$ $\propto T$ and 
$1/(T_{1}T) \propto$ $\langle D(\epsilon)\rangle_{kT}^{2} \propto T^{2}$, as in the case of 
quasi-particle excitations in $d$-wave superconductivity.
	The observed power law behaviors in shift and $1/T_{1}$ are consistent with this
zero-gap nature.
	The upward deviation of $1/(T_{1}T)$ from the $T^{2}$ dependence above 30 K may be related to the 
electron correlation effect or van-Hove singularity predicted to be located at a higher energy 
but remains to be seen \cite{Ref6_Kino}.

	As encountered in graphene with a similar band dispersion, impurity works as a dopant and 
easily shifts the Fermi level on the order of 10 meV \cite{Ref15_Novoselov} because of the vanishingly small 
density of states near the Dirac apex.
	For $\alpha$-I$_{3}$, the sign reversal of Hall coefficient is often observed 
at approximately 2 $\sim$ 7 K \cite{Ref16_Tajima} and is interpreted in terms of the passage of chemical potential 
across the Dirac apex \cite{Ref11_Kobayashi} as a manifestation of doping due to possible defects 
in the I$_{3}^{-}$ lattice.
	If the Fermi level is shifted from the cone apex, 
$\langle D(\epsilon_{\rm F})\rangle_{kT}$ should yield a finite 
value, giving $1/(T_{1}T)=$constant in the low-temperature limit.
	The persistence of $1/(T_{1}T) \propto T^{2}$ down to 10 K, at least in the present observation, 
means that the Fermi level shift is smaller than 1 meV.

	In conclusion, the magnetism of $\theta$-(BEDT-TTF)$_{2}$I$_{3}$, 
under pressure has been investigated by $^{13}$C NMR study.
	The linear temperature dependence of spin shift and 
the squared temperature dependence of $1/(T_{1}T)$ are both consistent with the picture 
that this material is a bulk zero-gap conductor.
%

	The authors thank N. Tajima, A. Kobayashi, S. Katayama, Y. Suzumura, A. Kobayashi, H. Kobayashi, 
and H. Fukuyama for helpful discussion.
	This work is partially supported by Grant-in-Aids for Scientific
Research on Priority Area (No. 17071003) and on Innovative Area (No. 20110002)
from MEXT, by Grant-in-Aids for Scientific Research (A) (No. 20244055) and (C) (No.
20540346) from the JSPS, and by the Global COE Program: Global Center of Excellence for the
Physical Sciences Frontier (No. G04).

\end{document}